# A REDUCED MOMENT MAGNETIC ORDERING IN A KONDO LATTICE COMPOUND: $Ce_8Pd_{24}Ga$

**D.T. Adroja, W. Kockelmann[†], A.D. Hillier, J.Y. So[*], K.S. Knight and B.D. Rainford[$]**

ISIS Facility, Rutherford Appleton Laboratory, Chilton, Didcot, Oxon, OX11, OQX, UK
[†]Mineralogisches Institut Universität Bonn, D-53115 Bonn, Germany
[*]Department of Physics, Seoul National University, Seoul 151-742, Korea
[$]Department of Physics and Astronomy, Southampton University, Southampton, SO17 1BJ, UK

The magnetic ground state of the antiferromagnet Kondo lattice compound $Ce_8Pd_{24}Ga$ has been investigated using neutron powder diffraction, inelastic neutron scattering and zero-field muon spin relaxation measurements. The neutron diffraction analysis, below $T_N$ (3.6±0.2K), reveals a commensurate type-C antiferromagnetic structure with the ordered state magnetic moment of ~0.36 $\mu_B$/Ce-atom along the cubic <111> direction. The analysis of the inelastic neutron scattering (INS) data based on the crystal field (CF) model reveals a doublet ground state with a ground state moment of 1.29 $\mu B$/Ce-atom. The observed magnetic moment from neutron diffraction, which is small compared to the expected value from CF-analysis, is attributed to screening of the local Ce moment by the Kondo effect. This is supported by the observed Kondo-type resistivity and a small change in the entropy of $Ce_8Pd_{24}Ga$ at $T_N$. The zero-field muon spin relaxation rate exhibits a sharp increase below $T_N$ indicating ordering of Ce moments, in agreement with the neutron diffraction data. The present studies reveal that the physical properties of $Ce_8Pd_{24}Ga$ are governed by the onsite Kondo compensation, the moment stabilizing intersite RKKY interaction and the crystal field effect.

# I. INTRODUCTION

Studies of Ce-based heavy fermion compounds are of fundamental importance for understanding their complex magnetic and transport properties. A number of recent discoveries have led to the resurgence of interest in the properties of the binary compounds which crystallise in the well known cubic $Cu_3Au$-type structure.[1-4] Among these, $CeIn_3$ and $CePd_3$ exhibit interesting properties. $CeIn_3$ shows antiferromagnetic ordering with $T_N = 10$ K at ambient pressure, and becomes superconducting with $T_C=0.175$ K at a pressure of 2.55 GPa.[1] On the other hand $CePd_3$ is a canonical intermediate valence compound which has been extensively investigated using many experimental techniques.[5-8] Despite of a low carrier density, 0.3 electrons per formula unit, the low temperature properties of $CePd_3$ are consistent with a simple Fermi-liquid ground state.[9] The commonality between $CeIn_3$ and $CePd_3$ is the presence of a broad peak in the resistivity, at 70 K and 150 K respectively, indicating the onset of a coherent Kondo lattice ground state.

In heavy fermion compounds the stability of the magnetic ordered state depends on a delicate balance between the long-range intermoment coupling driven by the Ruderman-Kittel-Kasuya-Yosida (RKKY) interaction and the on-site Kondo hybridization between the 4f electrons and the conduction electrons.[10,11] The strength of both the RKKY and Kondo interactions depends on the coupling parameter $(\rho J)$, $T_{RKKY} \sim (\rho J)^2$ and $T_K \sim \exp(-1/|\rho J|)$, where $J$ is the exchange parameter and $\rho$ is the density of states of the conduction electrons at the Fermi level. In the strong coupling limit, the so-called intermediate valence state as found in $CePd_3$, the ground state is non-magnetic. In this case both charge and spin fluctuations play an important role in the physical properties of the materials. For the medium to weak coupling limit, known as the heavy fermion regime, there can be a variety of interesting ground states, including reduced moment modulated antiferromagnetism, a zero temperature quantum critical point (QCP), unconventional superconductivity, and anomalous ferromagnetism. In this case the physical properties are governed by the spin fluctuations only. The coupling strength is generally related to the number of holes in the s-p or d-like outer shells of a given ligand and to its distance with respect to the Ce ions.[12,13] By decreasing the number of these holes it is possible to dehybridize the 4f states leading to a dramatic change in the magnetic and transport properties of the material. Such a dehybridization effect in $CePd_3$ has been observed when the compound was alloyed with an appropriate ligand or by introduction of small atoms such as B, Si or Ge at the interstitial site in the $Cu_3Au$ unit cell.[14-18] The interstitial impurities cause a dramatic cross-over from archetypal intermediate valence behaviour in $CePd_3$ to local moment character as seen through resistivity, magnetic susceptibility and inelastic neutron scattering studies.[14-18]

The crystal structure, magnetic and transport properties of $Ce_8Pd_{24}Ga$ have been investigated by various groups.[3,19-22] This compound is closely related to $CePd_3$ as the title compound may also be written $CePd_3Ga_{0.125}$. Our recent high resolution neutron powder diffraction study at 300 K on Ce$_8$Pd$_{24}$Ga revealed that the crystal structure is composed of distorted perovskite and Cu$_3$Au subcells arranged with the perovskite like units centred on the corners of the cube.[21] In the present work we have investigated the magnetic ground state of Ce$_8$Pd$_{24}$Ga using neutron powder diffraction and muon spin relaxation (μSR) measurements below and above T$_N$. The results of our inelastic neutron scattering data[21] are discussed in the context of the studies presented here.

## II. EXPERIMENTAL DETAILS

A polycrystalline sample (mass ~ 20 g) of Ce$_8$Pd$_{24}$Ga was prepared by arc melting the stoichiometric amounts of the constituent elements with purity 99.9% on a water-cooled copper hearth under high purity argon atmosphere. The ingot was turned over and remelted several times to ensure homogeneity. The neutron diffraction, inelastic neutron scattering and muon measurements were carried out at ISIS the pulsed neutron and muon source at Rutherford Appleton Laboratory, UK. Temperature dependent lattice parameters were measured between 4 K and 300 K using the high resolution neutron powder diffractometer HRPD. The magnetic structure was investigated as low as 1.9 K using the medium resolution neutron diffractometer ROTAX that views a cold methane moderator making the instrument suitable for magnetic structure studies. For both the ROTAX and HRPD diffraction experiments polycrystalline sample material was filled into a cylindrical vanadium sample container of 11 mm diameter (a flat plate type Al-sample holder was used for HRPD) and mounted in a standard 'orange-type' He-cryostat. Both HRPD and ROTAX are equipped with 3 detector banks, each of which provides neutron diffraction data in a distinct d-spacing and resolution range. Inelastic neutron scattering measurements were carried out using the time-of-flight spectrometer HET using an incident neutron energy of 40 meV. The Ce$_8$Pd$_{24}$Ga sample was cooled down to the lowest possible temperature of about 17 K in a closed-cycle refrigerator. Zero field μ$^+$SR measurements were carried on the MuSR spectrometer between 1.5 K and 6 K in a standard He-cryostat. The sample was mounted on an Al-sample holder with a silver mask. The neutron and μSR measurements were performed on the same sample of Ce$_8$Pd$_{24}$Ga.

## III. RESULTS AND DISCUSSION

### A. Neutron diffraction

The neutron powder diffraction studies show that down to the lowest temperature of 1.9 K Ce$_8$Pd$_{24}$Ga crystallises in the cubic structure with space group Pm-3m (no. 221), in agreement with our previous room temperature structurual analysis.[21] It is to be noted that the unit-cell of Ce$_8$Pd$_{24}$Ga is obtained by doubling the unit-cell of CePd$_3$ in all three cubic directions. The refinement of the site occupancies shows that the Ce and Pd sites are fully occupied, while the Ga site is only 78(4)% occupied. A detailed description of the crystal structure refinement and the crystallographic parameters at 300 K was given earlier.[21] Figure 1 shows the temperature dependence of the cubic lattice parameter obtained from Rietveld analysis. The lattice parameter decreases linearly with temperature from 300 K and becomes nearly temperature independent at low temperatures (below 15K), a typical behaviour expected for a system with thermally excited phonons. The temperature dependence of the lattice parameter was analysed using an Einstein expression of the form $a(T)=a_0+\kappa/[\exp(\vartheta_E/T)-1]$, where $a_0$ is the lattice parameter at 0 K, $\kappa$ is the Einstein constant, and $\vartheta_E$ is an effective Einstein temperature. The solid line in Figure 1 shows the resultant fit. The fitted parameters are: $a_0$=8.3669(1) Å, $\kappa$=0.0170(2) Å and $\vartheta_E$ =157(5) K. The Einstein model yields a reasonable fit to the data over the entire temperature range suggesting that the electronic contribution to the temperature dependence of the lattice parameter is small. We have estimated the Grüneisen parameter, $\gamma \sim 1.73$ for $Ce_8Pd_{24}Ga$, using the above values of $\kappa$ and $\vartheta_E$, and taking a typical value of the bulk modulus ~1Mbar.

Figure 2 shows the ROTAX forward scattering neutron powder diffraction pattern of $Ce_8Pd_{24}Ga$ collected at 1.9 K. At this temperature extra pure magnetic reflections with small intensities, such as (110) at a d-spacing of 5.95 Å, are observed in addition to the nuclear reflections. There are also additional small magnetic intensities observed on top of nuclear peaks, such as (211) (see inset Figure 2). These additional magnetic Bragg intensities are assumed to arise from the antiferromagnetic ordering of $Ce^{3+}$ moments at 1.9 K, in agreement with the heat capacity and resistivity measurements.[20,21] The observed magnetic reflections can be indexed on the basis of the crystallographic unit cell of $Ce_8Pd_{24}Ga$, i.e. with the wave vector q=(0, 0, 0). It is important to note that the same magnetic peaks can be indexed with respect to a smaller cubic $CePd_3$-type unit cell with a propagation vector [½, ½, 0]. This type of order was labelled type-C antiferromagnetic structure of a primitive cubic lattice earlier.[23] In the $Ce_8Pd_{24}Ga$ setting the Ce atoms occupy site 8g, site symmetry 3m, that leads to the admissible magnetic site symmetry 3m' restricting the Ce moments to directions along the 3-fold axis along <111>.[24] The corresponding magnetic space group is Pm3m'[25] corresponding to a triple-q structure model in terms of the [½, ½, 0] propagation vector, thus preserving the cubic symmetry. This non-collinear model, which has magnetic moments oriented along symmetry-equivalent <111> directions, represents the highest possible symmetry for the magnetic structure for fitting the magnetic intensities with one structure parameter only, i.e. the magnetic moment magnitude of the $Ce^{3+}$ ions. This magnetic structure cannot be discriminated by the diffraction experiments from other models of lower symmetry, e.g. a single-q structure. The restriction of the number of variable parameters was a main consideration for the model setup in view of the very weak observed magnetic powder diffraction intensities. The proposed structure takes into account the averaging of magnetic intensities over magnetic domains. The magnetic moment refinement of $Ce_8Pd_{24}Ga$ was performed with the program GSAS,[26] using diffraction patterns from the three banks in a Rietveld refinement. The magnetic form factor of $Ce^{3+}$ was calculated in dipole approximation using analytical approximations as tabulated in the International Tables for Crystallography.[27] The resultant Rietveld fit of the forward scattering bank data is shown in Figure 2. The refined moment value is 0.36(2) $\mu_B$/Ce-atom. A schematic representation of the magnetic structure of $Ce_8Pd_{24}Ga$ is displayed in Figure 3. The magnetic moment magnitude as a function of temperature is plotted in Figure 4. The magnetic moment of the Ce ion is saturated at low temperatures and almost constant at about 0.35 $\mu_B$/Ce-atom between 1.9 K and 3 K, rapidly dropping off to zero as the transition temperature is approached. Moment values of 0.2 $\mu_B$, corresponding to moment components of 0.1 $\mu_B$ along the cubic axes, are close to the detection limit of the powder diffraction experiment. Figure 4 also compares the temperature variation of the magnetic moment with a phenomenological power-law behaviour[28] $\mu(T) \sim (1 - \tau^d)^\beta$, where $\tau = T/T_N$ with $T_N=3.6$ K, where $\beta=0.43$ is taken from the muon spin relaxation measurements (see below) and where a value of d=10 ensures saturation of the magnetic moment curve below 3 K. The slightly higher value of $T_N = 3.6$ K obtained for the neutron diffraction sample of $Ce_8Pd_{24}Ga$ (mass ≈20 g), compared to $T_N=3.1$ K reported through the heat capacity and resistivity measurements (mass ≈5 g),[20,21] may be due to different Ga site occupancies in the different samples.

### B. Inelastic neutron scattering

Figure 5a shows the magnetic inelastic response from $Ce_8Pd_{24}Ga$ measured on HET at 17 K. The phonon scattering has been subtracted using a standard scaling method.[21,29] We observed two well-defined crystal field (CF) excitations at energies of 3.2 meV and 12.8 meV, corresponding to transitions from the ground state doublet to the two excited CF doublets. It is noteworthy that the inelastic response of $CePd_3$ does not show any sign of CF excitations, instead it exhibits a gap type response below 20 meV (i.e. no magnetic scattering) and a broad q-independent peak centred at 65 meV with a linewidth of 68 meV.[29,30] This is a typical response observed in valence-fluctuating systems, in which the strong hybridization between localised 4f electrons and conduction electrons is present.[31] The presence of well defined CF excitations in $Ce_8Pd_{24}Ga$ indicates that the strength of hybridization has reduced considerably. This is further supported by the dramatic reduction in Kondo temperature from 150 K for $CePd_3$ to 5.8 K for $Ce_8Pd_{24}Ga$. The point symmetry of the Ce site in $Ce_8Pd_{24}Ga$ is trigonal, 3m. This gives two additional terms in the cubic CF Hamiltonian of the $Ce^{3+}$ ion with J=5/2

$$H_{CF}=B_2^0 O_2^0 + B_4^0 O_4^0 + B_4^3 O_4^3 \qquad (1)$$

where $B_n^m$ are the crystal field parameters and $O_n^m$ are the Stevens operator equivalents.[32] $H_{CF}$ has only three independent CF parameters, thus the positions and relative intensities of the two CF excitations are sufficient to estimate a unique set of CF parameters. The values of the CF parameters obtained from the least-squares fit to the INS data are (in meV): $B_2^0=0.551(5)$, $B_4^0=0.024(2)$, and $B_4^3=0.208(4)$. The solid line in Figure 5a represents the fitted response with this set of CF parameters, which agrees well with the observed magnetic response. Figure 5b represents the crystal-electric-field energy level scheme of the $Ce^{3+}$ ions in $Ce_8Pd_{24}Ga$ with the above set of CF parameters. The ground state is $|\pm3/2\rangle$, which makes $Ce_8Pd_{24}Ga$ an Ising system. The calculated value of the ground state magnetic moment, $\langle\psi_1^{\pm}|g_JJ_z|\psi_1^{\pm}\rangle$ (where $g_J$ is the Landé g-factor), is 1.29 $\mu_B$. By applying the CEF theory in the lowest order approximation, the magnetocrystalline anisotropy energy $E_a$ can be expressed in term of the second-order CEF parameter $B_2^0$ as follows

$$E_a = -3/2\, B_2^0\, \langle 3J_z^2 - J(J+1)\rangle \sin^2\theta \qquad (2)$$

where $\theta$ is the angle between the easy magnetization direction and the z-axis (i.e. the crystal electric field axis, which is <111> in the present case). The observed positive sign of $B_2^0$ indicates that the magnetic moment is along the <111> direction in $Ce_8Pd_{24}Ga$, which is in agreement with the neutron diffraction analysis.

### C. Zero field muon spin relaxation

The zero field muon spin relaxation is a unique technique and has played an important role to investigate dynamical fluctuations and static inhomogeneities in reduced moment heavy fermion systems.[33] In order to shed more light on the nature of the magnetic ground state in $Ce_8Pd_{24}Ga$, zero field μSR measurements have been carried out on a polycrystalline sample below and above $T_N$. Figure. 6 shows zero field μSR spectra from $Ce_8Pd_{24}Ga$ measured at various temperatures. The zero field μSR spectra show a very small depolarization rate in the paramagnetic state, while below 3.8 K an onset of strong muon depolarization is observed, which is in agreement with the antiferromagnetic ordering observed by neutron diffraction. Surprisingly, even well below the magnetic ordering temperature there is no clear sign of a coherent oscillation in the zero field μSR spectra that would give evidence of a muon spin precession around a static internal magnetic field. A very similar situation has been observed for the antiferromagnet $YbCu_3Al_2$ ($T_N$=2 K) and ferromagnetic CePdSb ($T_C$=17.5 K).[34,35] This might suggest that the average dipolar field, experienced by the muon spins at the muon stopping site, is equal to zero in $Ce_8Pd_{24}Ga$. In this scenario the sharp increase of the muon depolarization rate observed below $T_N$ in $Ce_8Pd_{24}Ga$ could be understood as a result of the distribution of internal fields with zero mean at the muon site.

In the paramagnetic state the zero field μSR spectra are well described using a product of a temperature-independent Kubo-Toyabe (KT) function, arising from the Ga and Pd nuclear moments, and an exponential relaxing function (ERF), $\exp(-\lambda_1 t)$, representing the additional muon depolarization due to fluctuating electronic 4f-magnetic moments. The dynamic relaxation rate, $\lambda_1$, is related to the interaction amplitude and to the rate of the spin fluctuations. In our analysis, we have also added a small time and temperature independent background contribution ($a_{bg}$=0.026, estimated from the spectra at 1.6 K) arising from muon stopping in the silver mask. Fits to data collected at temperatures above $T_N$ yield good agreement between observed and calculated spectra as shown in Figure 6. The estimated KT relaxation rate is $\sigma_{KT}$=0.0306 $\mu s^{-1}$. The comparison of this value for $\sigma_{KT}$ with the calculated value from the second moment of the internal nuclear dipole field distribution across the unit-cell of $Ce_8Pd_{24}Ga$ indicates that the muons are localised at the (0, ¼, ¼) site. However, the above mentioned functional form does not fit the data collected below $T_N$. Data modelling with various forms of alternative, single component muon spin relaxation functions did not result in satisfactory fits of the data below $T_N$. The spectra could be best fitted over the entire temperature range by adding a second component to the product of KT and ERF for the temperature regime below $T_N$:

$$G_z(T) = G_{KT}(\sigma_{KT}, t) [\ a_1 \exp(-\lambda_1 t) + a_2 (1/3+2/3 \exp(-(\lambda_2 t)^2))\ ] + a_{bg} \qquad (3)$$

where $G_{KT}(\sigma_{KT}, t)$ is the well-known Kubo-Toyabe function[33] and the asymmetry parameter $a_2$ preceding the additonal second term. The first and third terms are also used for fitting above $T_N$ whereas the second term contributes below $T_N$ only. The functional form of the second term has been suggested for a muon stopping site of high symmetry in an antiferromagnetic matrix.[36] At such a site the mean field is zero. From the neutron powder diffraction experiments it follows that 22(4)% of the Ga site is vacant in $Ce_8Pd_{24}Ga$, hence one expect a different value of moment for the Ce atoms around this vacant site, which justifies to use the second term and, hence, the two components model in our analysis. In the analysis, the values of total asymmetry, a=a1+a2, $a_{bg}$ and $\sigma_{KT}$ were kept fixed. The temperature dependent relaxation rates, $\lambda_1$ and $\lambda_2$, and asymmetries, $a_1$ and $a_2$, obtained from the fits are displayed in Figure 7a and 7b. Below $T_N$ $\lambda_2$ increases sharply on cooling, while $\lambda_1$ shows a very weak temperature dependence (Figure 7a). On the other hand, the asymmetry $a_2$ decreases sharply when the temperature is raised towards $T_N$, accompanied by a compensation rise of $a_1$ (Figure 7b). These types of two components in the muon depolarization rate have been also observed in antiferromagnetic $CeAl_2$ and the heavy fermion YbBiPt.[33] The presence of two components has been attributed to two types of domains in these materials. In $Ce_8Pd_{24}Ga$ the fast depolarization rate ($\lambda_2$) is associated with the magnetic ordered domain, while the slow depolarization rate ($\lambda_1$) is associated with the paramagnetic domain. The asymetry (or amplitudes) of the fast component ($a_2$) represents the magnetic volume fraction, and the asymmetry of the slowly damped component ($a_1$) reflects the paramagnetic volume fraction. The volume fractions of magnetic and paramagnetic domains at 3.1 K (1.6 K) are 77% (88%) and 23% (12%), respectively. These values agree with the fractional occupancy of the Ga site (78%) obtained from the high resolution neutron powder diffraction measurements. This indicates that the Ce ions adjacent to Ga vacancies remain paramagnetic below $T_N$.

The temperature dependence of $\lambda_2$ has been modelled using the same functional power-law term that was plotted with the ordered magnetic moment in Figure 4, namely $\lambda_2(T) \sim (1 - \tau^d)^\beta$ (Figure 7a), yielding $T_N$=3.51(1) K, $\beta$=0.429(1) and d=2.05(2). The value of $T_N$ obtained from the muon data is in agreement with that obtained from the neutron diffraction study. The phenomenological parameter of d=2.05, reflecting spin-wave excitations,[28] is very close to 2 predicted for a cubic antiferromagnetic system.[37] We do not clearly understand at present the disagreement with the much higher value of d=10 employed to obtain a saturated ordered magnetic moment below 3 K (Figure 4). One of the reasons could be a few data points with a large errors near $T_N$ in the diffraction data.

## IV. CONCLUSIONS

In the present study, we have investigated the antiferromagnetic Kondo lattice compound $Ce_8Pd_{24}Ga$ using neutron diffraction, inelastic neutron scattering and zero-field muon spin relaxation measurements. The chemical formula of the title compound can also be written as $CePd_3Ga_{0.125}$ in order to make a direct comparison with $CePd_3$. The present study reveals that the physical properties of $CePd_3$ change strongly with the incorporation of the Ga atom into the cubic cell. The change of the physical properties could be attributed to the dehybridization of Ce-4f and Pd-4d electrons as a result of the shorter bond-length (2.453 Å) between Pd-Ga atom, that produces a strong overlap between the p-electrons of the Ga and d-electrons of the Pd, and hence the strong dehybridization in $Ce_8Pd_{24}Ga$. The effect of dehybridization is clearly seen through the drastic change in the Kondo temperature ($T_K$), 150 K for $CePd_3$ and 5.8 K for $Ce_8Pd_{24}Ga$.[21] The observation of the magnetic ordering in $Ce_8Pd_{24}Ga$ is consistent with the prediction by Doniach's phase diagram, which reveals that with reducing the hybridization (or the coupling constant, $\rho J$), the Kondo temperature decreases and the magnetic ordering temperature increases. Furthermore, the observation of well defined crystal field excitations in $Ce_8Pd_{24}Ga$ indicates that the Ce ions are in the localized state.

The neutron diffraction study of $Ce_8Pd_{24}Ga$ shows that the Ce magnetic moment in the order state is ~0.36 $\mu_B$/Ce-atom, compared with the predicted CF ground state moment 1.29 $\mu_B$/Ce-atom from the crystal field eigenstate, which clearly indicates the reduced moment magnetic ordering. The study of reduced moment magnetic ordering in Ce, Yb and U-based heavy fermion compounds is an active area of research in the field of heavy fermion systems. This type of reduced moment antiferromagnetic ordering has been observed in $Ce(Ni_{0.65}Pd_{0.25})_2Ge_2$, with an estimated ordered state moment of 0.55 $\mu_B$/Ce-atom, and in the heavy fermion antiferromagnetic superconductor $UPt_3$ ($T_c$=0.5 K and $T_N$=5 K), with an estimated ordered state moment of 0.02 $\mu_B$/U-atom.[38,39] The presence of the Kondo effect in $Ce_8Pd_{24}Ga$ has been confirmed through the temperature dependence resistivity, the scaling behaviour of the magnetoresistance, and the observed small entropy change, 16.4 J mole$^{-1}$ K$^{-1}$ compared with the 8RLn(2)=46.1 J mole$^{-1}$ K$^{-1}$ expected from a doublet CF ground state.[20,21] Therefore, the origin of the reduced moment ordering in $Ce_8Pd_{24}Ga$ is attributed to the screening of the local 4f-moment by the conduction electrons by forming a Kondo-singlet state. The energy gain due to the singlet formation is $K_B T_K$. It should be noted that below 8 K $Ce_8Pd_{24}Ga$ shows an onset of the coherence in the resistivity, i.e. Kondo lattice behaviour, implying the importance of interactions between the magnetic ions of the lattice. In this scenario, while gaining the energy through the singlet formation, the system is also losing the magnetic interaction energy between the sites. This implies that the singlet formation should not be fully complete when the intersite interaction is sufficiently strong, and hence the system orders magnetically with a reduced magnetic moment. This is because the fully Kondo compensation of the 4f-moments would lead to an overall loss of energy instead of an energy gain. Thus, the Kondo compensation model explains the reduced moment magnetic ordering in $Ce_8Pd_{24}Ga$.

## Acknowledgements

The authors gratefully acknowledge valuable discussions with A.T. Boothroyd, Physics Department Oxford University and B.FäK, ISIS facility, Rutherford Appleton Laboratory. One of us (W.K.) acknowledges financial support by the German Bundesminister für Bildung und Forschung, project 03KLE8BN.

**Figure captions**

Figure 1:  Temperature variation of the lattice parameter of $Ce_8Pd_{24}Ga$ which has been modelled using an Einstein expression (solid line).

Figure 2:  Forward scattering neutron diffraction pattern of $Ce_8Pd_{24}Ga$ collected on ROTAX at 1.9 K, with observed (symbols), calculated (line) and difference patterns (lower curve) obtained by a Rietveld refinement of nuclear and magnetic structure parameters. The inset shows the enlarged diffraction in a d-spacing range between 3-7 Å. N and M mark the positions of nuclear and magnetic Bragg intensities, respectively.

Figure 3:  Schematic of the proposed magnetic structure of $Ce_8Pd_{24}Ga$. The magnetic moments vectors are aligned along <111> directions. Ce moments in [110] and [1–10] sheets point into the upper and lower corners of the unit cell, respectively, in line with the C-type structure and a [½, ½, 0] propagation direction with respect to a $CePd_3$-type cell.

Figure 4:  Temperature dependence of magnetic Ce moment in $Ce_8Pd_{24}Ga$, obtained from the Rietveld analysis of the neutron diffraction data. The solid line represents a power-law behaviour with $T_N$=3.6 K and $\beta$=0.42 (see text).

Figure 5a:  Magnetic response from $Ce_8Pd_{24}Ga$ at 17 K (symbols) measured on the HET spectrometer, the data summed over the scattering angles between 8° and 30°. The solid line represents the fit based on the crystal field model (see text).

Figure 5b:  Crystal field energy level scheme of $Ce_8Pd_{24}Ga$ with eigenstates.

Figure 6:  Zero-field μSR spectra of $Ce_8Pd_{24}Ga$ at temperatures above and below $T_N$. The solid lines represent fits to the data (see text). The inset shows the spectra at small time for 1.6K.

Figure 7a:  Temperature dependence of the zero-field muon spin relaxation rate, $\lambda_1$ and $\lambda_2$. The solid line represents a power-law fit (see text). The inset shows $\lambda_1$ versus temperature plotted with an expanded y-scale.

Figure 7b:  Temperature dependent asymmetry of two components. Note that the total asymmetry was kept fixed during the analysis. The dotted line lines are guid to the eye.

Figure 1

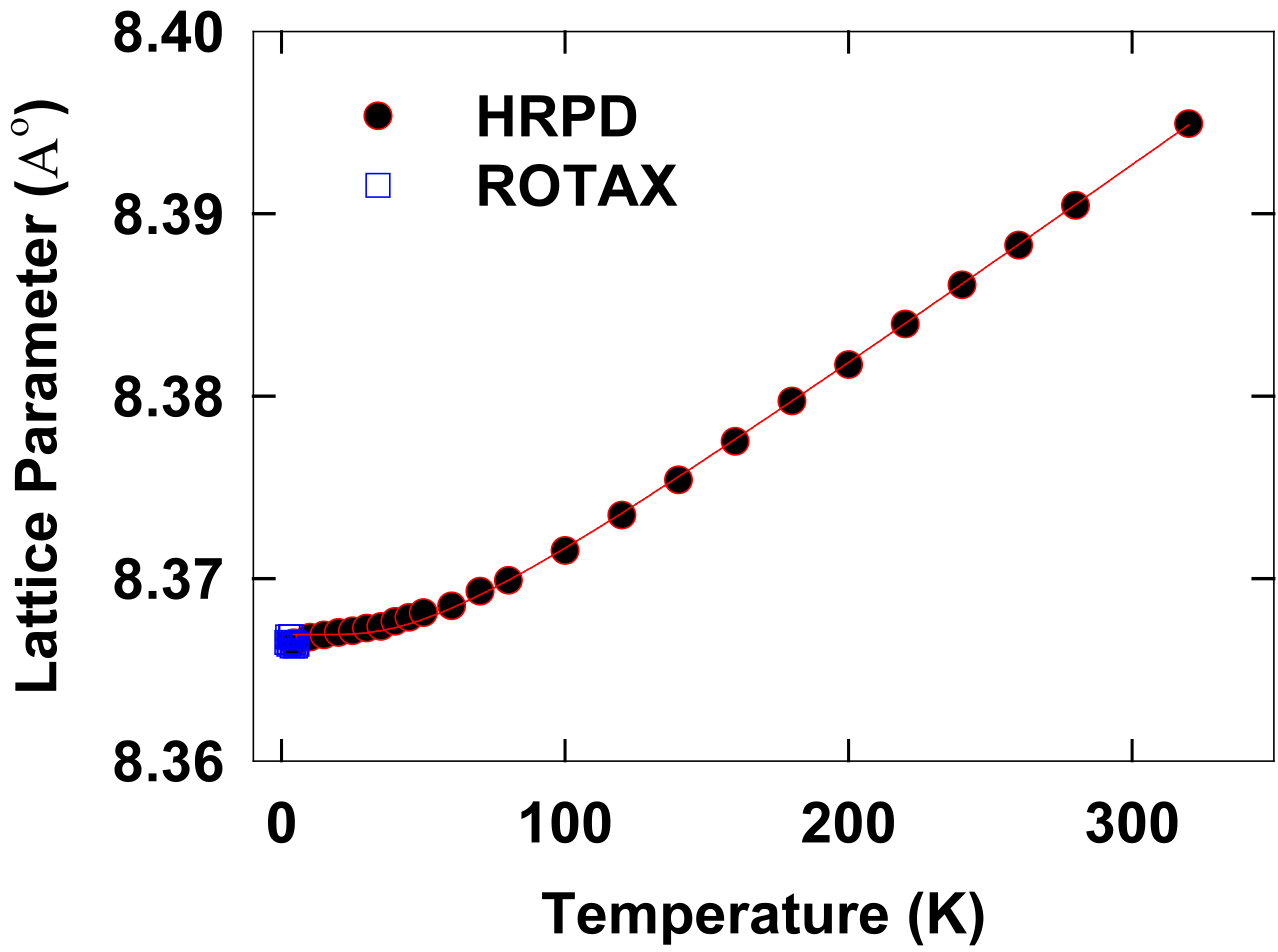

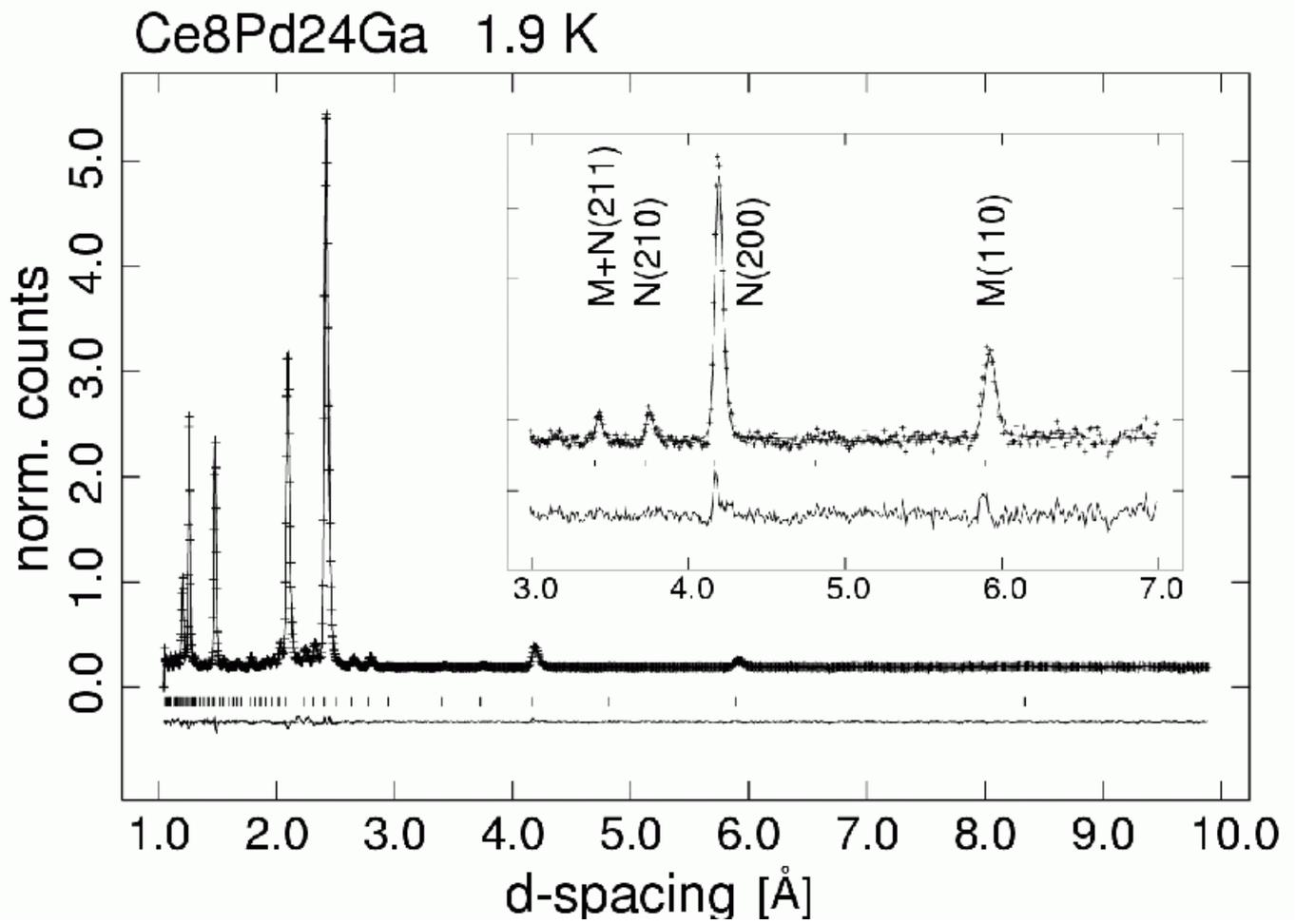

Figure 3

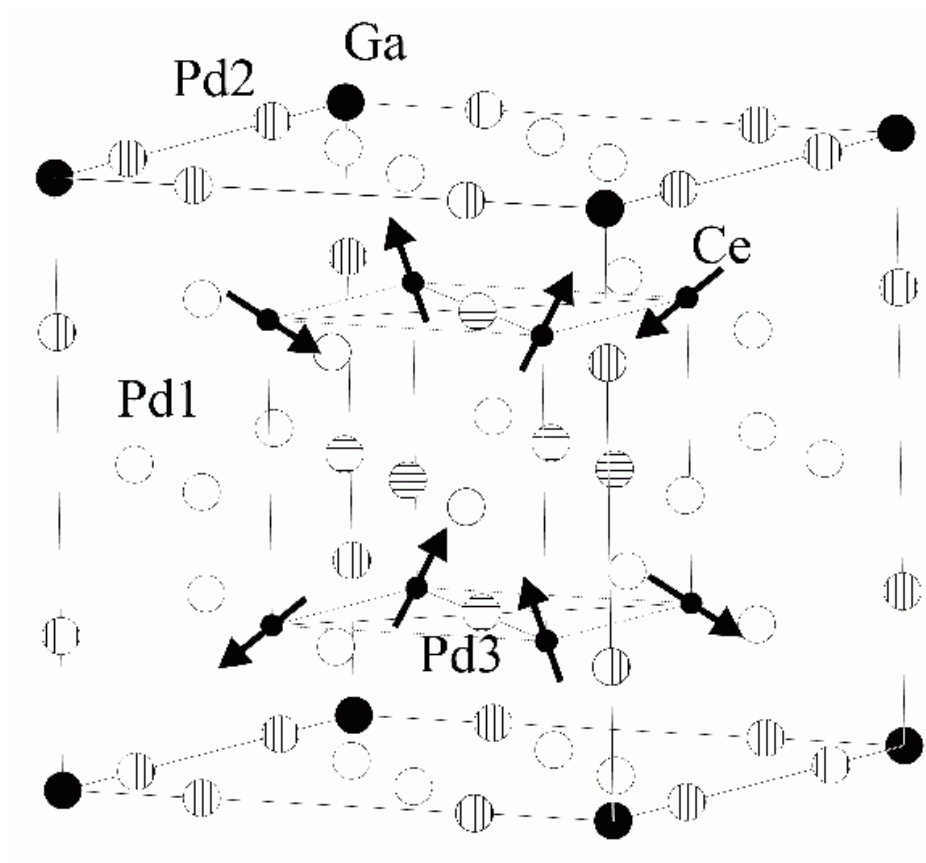

Figure 4

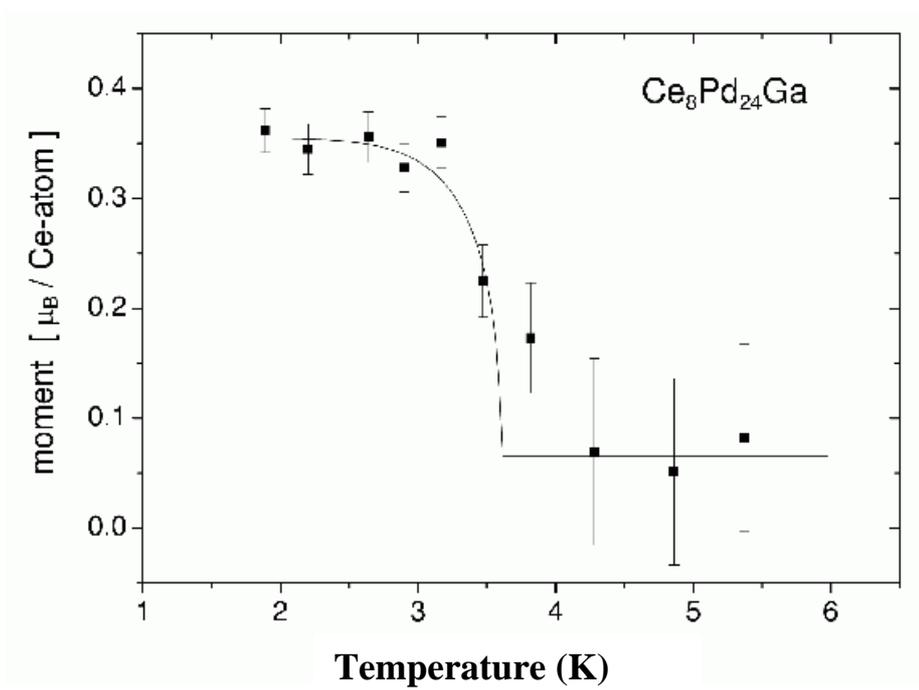

Figure 5a

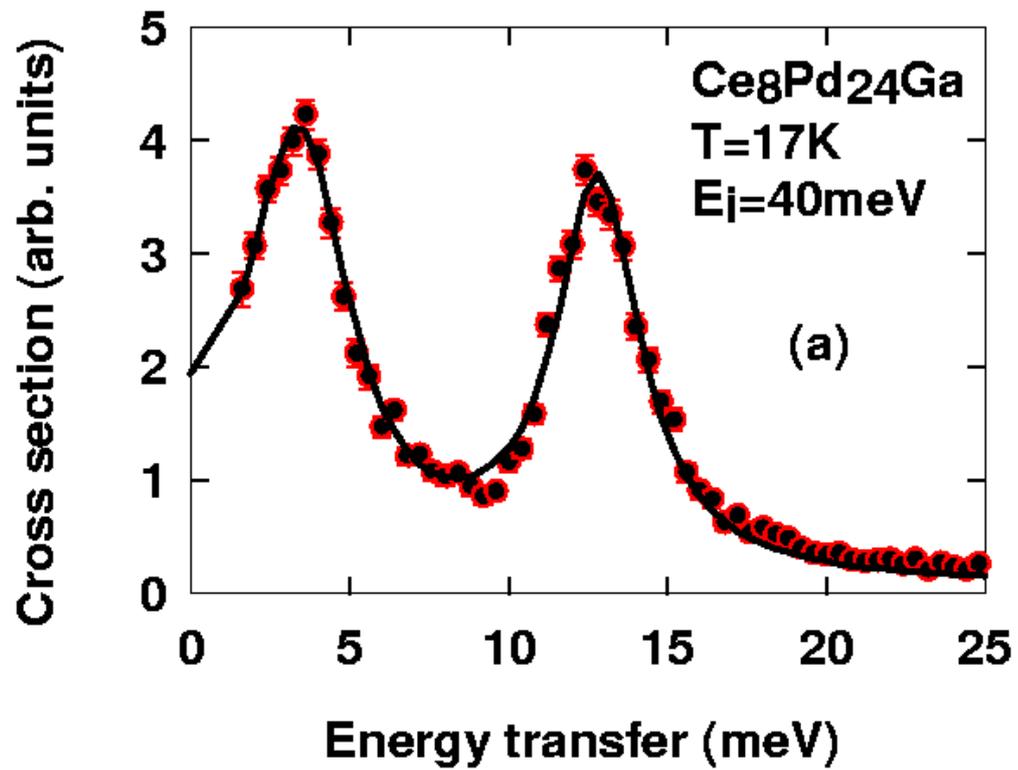

Figure 5b

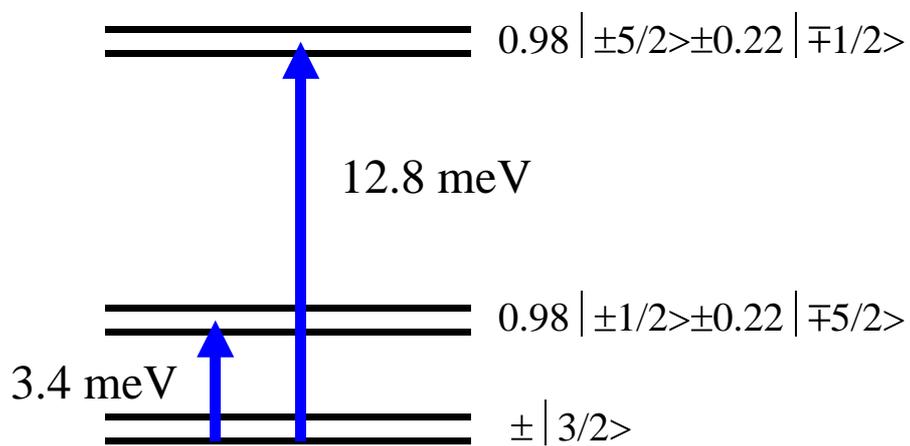

Figure 6

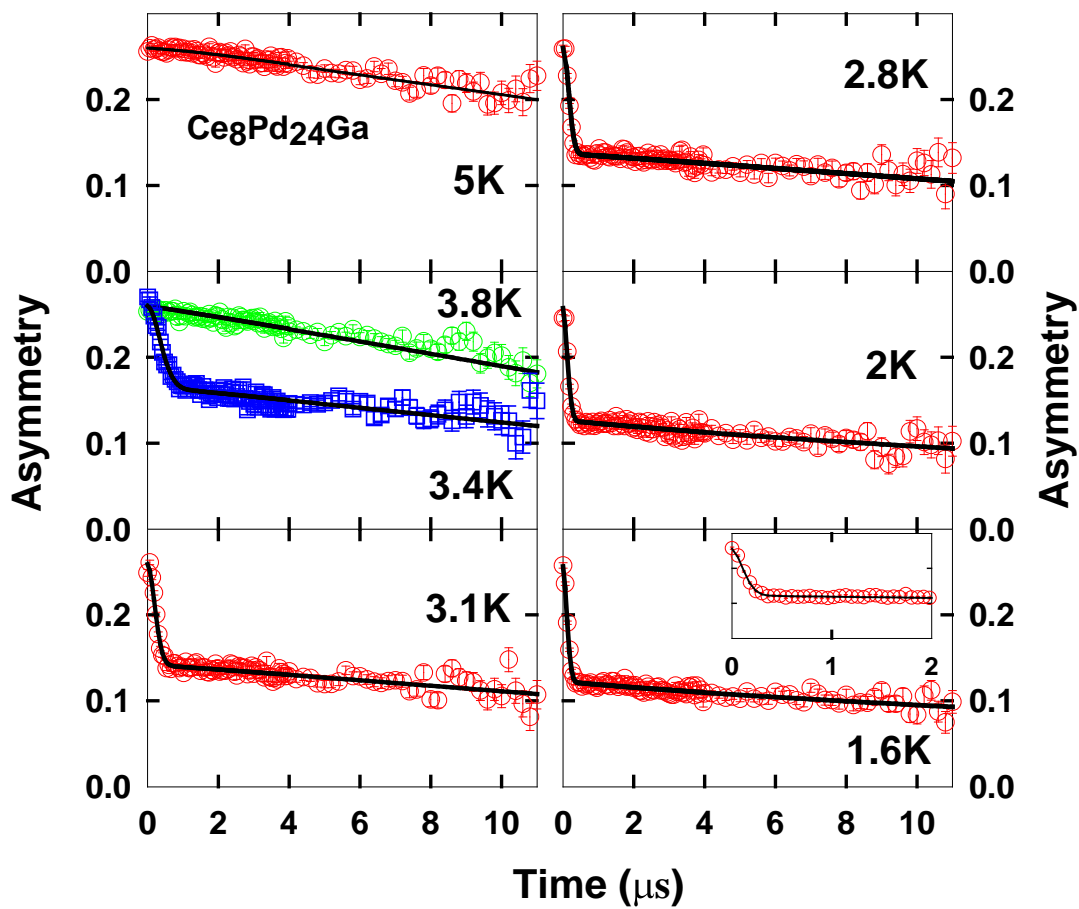

Figure 7a and 7b

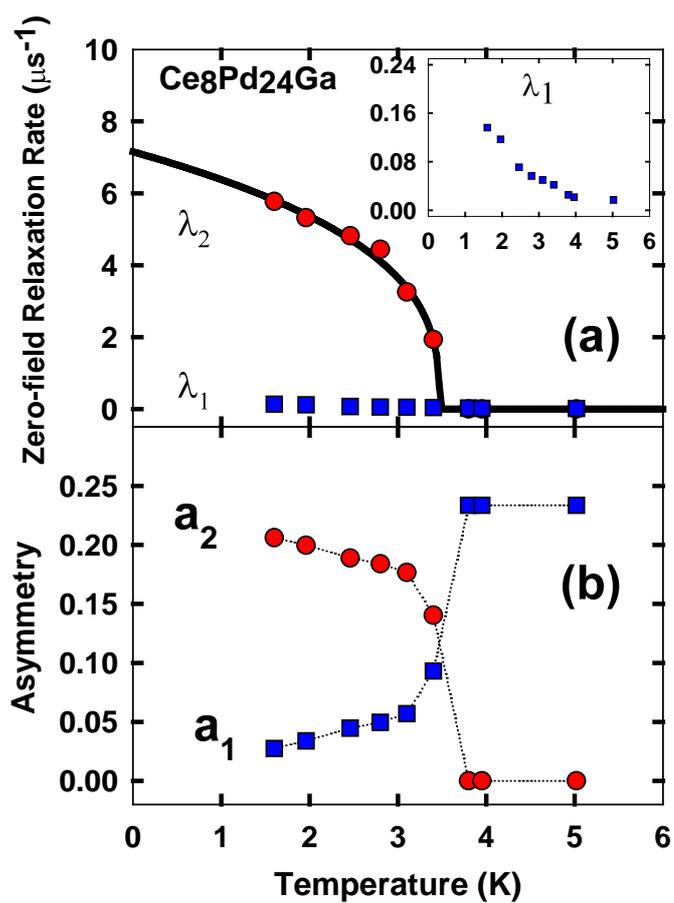